\input phyzzx
\hoffset=0.375in
\overfullrule=0pt
\def\bs{{\bf s}}
\def\bx{{\bf u}}
\def\br{{\bf r}}
\def\bl{{\bf l}}

\def\mathbf{\bf}
\def\mathrm{\rm}
\font\bigfont=cmr17
\centerline{\bigfont
Finite Source Effects in Microlensing Events}
\bigskip
\centerline{{\bf Andrew Gould}\footnote{1}{Alfred P.\ Sloan Foundation Fellow}}
\smallskip
\centerline{Department of Astronomy, Ohio State University}
\smallskip
\centerline{Columbus, OH 43210}
\bigskip
\centerline{and}
\bigskip
\centerline{{\bf C\'edric Gaucherel}}
\smallskip
\centerline{Centre d'Etudes de Saclay, 91191 Gif-sur-Yvette, France}
\bigskip
\centerline{gould@payne.mps.ohio-state.edu, gauche@hep.saclay.cea.fr}
\bigskip
\centerline{\bf Abstract}
\doublespace
\singlespace

The computation of the magnification of a finite source by an arbitrary 
gravitational lens can be reduced from a two-dimensional to a one-dimensional 
integral using a generalization of Stoke's theorem.  For a large source lensed
by a planetary-system whose planet lies at the position where one of the two 
images would be in the absence of a planet, the integral can be done 
analytically.  If the planet lies at the position of the major (unperturbed) 
image, the excess flux is the same as it would be for an isolated planet.  If 
the planet lies at the minor image, there is no excess flux.

\bigskip
\noindent Subject Headings: gravitational lensing -- methods: numerical -- 
planetary systems
\smallskip
\noindent Submitted to {\it The Astrophysical Journal Letters} June 17, 1996
\smallskip
\noindent Preprint: OSU-TA-17/96
\endpage

\normalspace

\chapter{Introduction} 

	Four groups have detected more than 100 microlensing events toward
the Large Magellanic Cloud and the Galactic bulge 
(Alcock et al.\ 1995,1996a; Aubourg et al.\ 1995; Udalski et al.\ 1994a;
Alard 1996).  For most events, the source can be treated as point of
light.  However, when the source comes sufficiently close to or crosses a 
caustic (locus of points of infinite magnification in the lens plane), the 
finite size of
the source affects the light curve.  One may use these effects to infer
the size of the Einstein ring relative to the angular size of the source.
Since the latter is generally known from Stefan's law and the color and 
magnitude of the source,
one can then determine the absolute size of the Einstein ring (Gould
1994; Nemiroff \& Wickramasinghe 1994).  This effect has already been observed
for one point-mass lens (Pratt et al.\ 1996) and for two binary lenses
(Udalski et al.\ 1994b; Alcock et al.\ 1996b,1996c), and may ultimately
be key to measuring the mass function of the lenses (Gould 1996).

	For a point-mass lens, one may write the formula for the magnification
of a finite source in closed form (Witt \& Mao 1994), but for a binary
lens, the evaluation is more complicated.  In principle, one could 
compute the magnification at each point of the source and sum these to find
the total flux of the images.  However, because the magnification is divergent
near the caustic, one must take special care in performing the integration
in these regions.  Since the caustics have a somewhat irregular structure, this
form of numerical integration is often difficult.

	The problem can be especially acute in the analysis of lensing events
by planetary systems because the Einstein ring of a planet is generally of the
same order as the size of the source.  In order to simulate such events 
Bennett \& Rhie (1996) developed an alternate approach:  they examined the 
points in the image plane (rather than the source plane), 
calculated the source-plane 
position for each, and thereby identified all the image points originating
in the source.  For a source of uniform surface brightness, this method 
yields the ratio of the total area of the images to the area of the source 
which, since surface brightness is conserved (Liouville 1837), is equal to 
the total magnification.  The method is easily generalized to non-uniform
sources by weighting each point of the image by the local flux of the
corresponding point on the source.
Here we present a new method for computing the magnification of finite sources.

\chapter{Method}

	Initially we will assume 
that the source has uniform surface brightness so that by Liouville's theorem
the magnification is just the ratio of the area of the image to the area of
the source.  Later we will extend the method to more general sources.

	Consider first a source that does not cross any caustics.  
The source will be imaged into $m$ disjoint images.  Let $C$ be the boundary 
of the source and let $C_j'$ be the boundary of the $j$th image.  The parity
of each image, $p_j=\pm 1$, is defined as the sign of its magnification tensor.
As one moves counter-clockwise around $C$, one moves counter-clockwise around
$C_j'$ for $p_j=1$ and clockwise for $p_j=-1$.  By Stoke's theorem, the
area of the source is $(1/2)\int_C \br \times d\bl$ and the area of the
$j$th image is $(1/2)p_j \int_{C_j'}\br \times d\bl$,
where $\br$ is the position on the contour and $d\bl$ is the line element.
Note that the direction of integration around the image contours is defined
by counter-clockwise motion around the source.  The magnification is then
$$A = \sum_j p_j \int_{C_j'}\br \times d\bl\bigg/\int_C \br \times d\bl,
\eqn\stokesone$$
where the two-dimensional cross products are to be regarded as signed scalars.

	Equation \stokesone\ remains valid even when the source crosses one or
several 
caustics.  To see this, divide the source into subsources each of which lies
entirely inside or entirely outside of caustics.  For definiteness, take
the case of a binary lens for which the source can be divided into 
two subsources, one lying inside a caustic and having five images
and the other lying outside and having three images.  The magnification is
then given by the sum of two integrals of the form of equation \stokesone,
one integral for each subsource.
The difference between this sum and equation \stokesone\ applied directly
to the whole source is eight additional line integrals, five for the
image contours mapped from motion in one direction along the inside of the 
caustic segment, and three for the image contours mapped from motion in the
opposite direction along the outside of the caustic segment.  Consider first
the two images that are present inside but not outside the caustic.  These
have opposite parities and, for points along the caustic, are mapped into
exactly the same points along the critical curve in the image plane.  
Hence the two line integrals
from these images make equal contributions of opposite sign.  Now consider
the remaining three images.  These are unaffected by the presence of the
caustic and therefore the contours just inside and just outside the caustic
are mapped to the same contours in the image plane.  However, since the
directions of integration are opposite, the two line integrals cancel for
each image.  Thus, equation \stokesone\ is valid for all cases.

\chapter{Application to Planetary Systems}

	Consider a planet of mass $m$ orbiting a star of mass $M$, with
$m\ll M$.  If the planet were not there, the star would lens a background
source into two images at positions $\pm x_\pm\theta_e$ where $\theta_e$ is the
angular Einstein radius of the lensing star,
$$x_\pm \equiv {(x^2+4)^{1/2} \pm x\over 2},\eqn\xpmeq$$
and $x\theta_e$ is the projected separation between the source and the lens.
The magnification tensor is given by
$${\cal M}_\pm = \left(\matrix{1+\gamma_\pm & 0\cr 0 & 1-\gamma_\pm}
\right)^{-1},
\qquad \gamma_\pm = x_\mp^2,
\eqn\calm$$
where the (1,1) element represents the magnification along the source-lens
axis.  The magnification of each image is given by the absolute value of
the determinant of this tensor, $A_\pm=|{\cal M}_\pm|$.
Note that the shear $\gamma_+<1$ for the major image outside the
Einstein ring ($x_+>1$) and that
$\gamma_-=\gamma_+^{-1}>1$ for the minor image inside the Einstein ring
($x_-<1$).

	We now suppose that the planet lies exactly at the position of one
of the two {\it unperturbed} images of the center of the source.  We adopt
this position as the center of our coordinates and express all angular
distances in
units of the Einstein ring of the {planet}: $\theta_p=(m/M)^{1/2}\theta_e$.
We denote positions within the source by $(\rho\cos\psi,\rho\sin\psi)$ and
positions within the image by $(r\cos\phi,r\sin\phi)$.  We evaluate
equation (2.14) from Gould \& Loeb (1992), noting that in their notation
$(\rho\cos\psi,\rho\sin\psi)=
-\epsilon^{-1/2}([1+\gamma]\xi_p,[1-\gamma]\eta_p)$
and
$(r\cos\phi,r\sin\phi)=\epsilon^{-1/2}(\xi_i-\xi_p,\eta_i-\eta_p)$.
We then find,
$$\rho\cos\psi = {\cos\phi\over r}[r^2(1+\gamma)-1],\qquad
\rho\sin\psi = {\sin\phi\over r}[r^2(1-\gamma)-1].\eqn\glcoords$$
Squaring and adding these two equations yields a quadratic equation in $r^2$,
the two solutions of which are
$$r_\pm^2 = {b\pm (b^2 - 4 a)^{1/2}\over 2 a},\quad 
a\equiv 1+\gamma^2+2\gamma\cos2\phi,\quad b\equiv \rho^2+2+2\gamma\cos2\phi.
\eqn\rpmeval$$

	Suppose that the source is large enough so that it covers all
caustics (see e.g.\ fig.\ 3 from Gould \& Loeb 1992).  The boundary of the
source will then always have two images, one at $r_+$ and one at $r_-$.
Using equation \stokesone, and assuming that the source has constant 
surface brightness, we find a magnification 
$$A = {1\over 2\pi \rho^2}\int_0^{2\pi}d\phi\,(r_+^2-r_-^2) 
= {1\over |1-\gamma^2|} + {1+{\rm sgn}(1-\gamma)\over\rho^2} -
{q\over \rho^4} + ...
.\eqn\aeval$$
where 
sgn$(1-\gamma_\pm)=\pm 1$ and $q=[(1/2+\rho^{-2})^2-\gamma^2\rho^{-4}]^{-1/2}$.
The first term is just the magnification
of the source in the absence of a planet [cf.\ eq.\ \calm].  
For the major image, the second
term is $2\rho^{-2}$.  Thus, for a source of unit surface brightness
the total excess flux is $2\pi\theta_p^2$, exactly the same as the result for
an isolated planet.  On the other hand, to this order there is no excess flux
when a planet perturbs the minor image, a result already suggested by
the numerical calculations of Bennett \& Rhie (1996).  Successive additional
terms are each smaller by $\rho^{-2}$.

\chapter{Numerical Integration}

	To translate equation \stokesone\ into a prescription for numerical
integration, we first approximate the boundary of the source as a polygon of 
$n$ (not necessarily equal) sides.  We denote the (two-component) vertices in 
counter-clockwise order by $\bs_0$, $\bs_1$ ... $\bs_n$, with $\bs_n=\bs_0$. 
For each source vertex $\bs_i$, there will be a variable number of image 
positions $\bx_{i,j}$.  The vertex images should be 
ordered so that $\bx_{i-1,j}$ and $\bx_{i,j}$ lie on the same image curve.
When the source contour crosses a caustic and two images disappear, 
these images should be replaced by ``blanks''.  When a caustic is crossed
and two new images appear, they should be entered into previously blank
columns.  With this ordering, the parities of the image vertices depend only
on $j$: $p_{i,j}\rightarrow p_j$.
For simplicity, we initially assume that if any caustics are crossed, one of 
the source vertices is chosen to lie right on the caustic.  

	The magnification is then given by,
$$A = \sum_{i=1}^n \sum_{j'} p_j(\bx_{i-1,j}\times\bx_{i,j})\bigg/
\sum_{i=1}^n \bs_{i-1}\times\bs_{i},\eqn\mageq$$
where the prime in $j'$ indicates that there is no summation for first
appearance of new images at a caustic (in which case there is, of course,
no previous image position $\bx_{i-1,j}$).

	In equation \mageq, we assumed that if the source boundary crossed
a caustic (moving counter-clockwise) thereby created or destroying two
images, then one of the vertices would be chosen to lie exactly on the
caustic.  We now show that if the first point does not lie on the caustic,
there is a simple prescription which in effect replaces the two terms 
connecting the critical curve and the two images of the first point inside
the caustic with a single term that connects the two image points directly.
Let $j$ and $j+1$ be two new images and let $\bs_{i}$ be chosen to 
lie exactly on the caustic where they are created.  
The first term to be included in the sum for 
the $j$ image would be $i+1$ and this term would include the boundary 
between the critical curve (at $\bx_{i,j}$) and the point at $\bx_{i+1,j}$.  
The situation is similar for image $j+1$.  The two new images have opposite
parities, $p_{j+1}=-p_j$.  Because $\bs_i$ lies on the caustic, 
$\bx_{i,j}=\bx_{i,j+1}$.
The sum of the first terms for these two new images will then be
$$p_j(\bx_{i,j}\times \bx_{i+1,j})+p_{j+1}(\bx_{i,j+1}\times \bx_{i+1,j+1})
= p_j \bx_{i,j}\times (\bx_{i+1,j}-\bx_{i+1,j+1})
.\eqn\exct$$
To a good approximation $\bx_{i,j}=(\bx_{i+1,j}+\bx_{i+1,j+1})/2$, so 
one may simply replace the two terms on the left-hand side of equation 
\exct\ with $p_j \bx_{i+1,j+1}\times\bx_{i+1,j}$.  Now let $\bs_{i'}$ be the
vertex on a caustic where the two images disappear.  Using a similar argument, 
one can show that the two last terms for these images can be replaced by
$-p_j \bx_{i'-1,j+1}\times\bx_{i'-1,j}$.
Hence, it is not actually necessary to have vertices on the caustics. 
Suppose that there are $k$ caustic crossings, $\ell=1 ... k$ where two images
$j_\ell$ and $j_\ell+1$ are created, and $k$ other crossings where they
are destroyed.  Let the first point after the images have been created be
$i_\ell$ and last before they are destroyed be $i'_\ell$.  If these first
and last points do not lie on the caustic, then the numerator in equation
\mageq\ should be replaced by

$$ 
\rightarrow
\sum_{i=1}^n \sum_{j'} p_j(\bx_{i-1,j}\times\bx_{i,j})
+\sum_{\ell=1}^k p_{j_\ell}[\bx_{i_\ell,j_\ell+1}\times\bx_{i_\ell,j_\ell}
-\bx_{i'_\ell,j_\ell+1}\times\bx_{i'_\ell,j_\ell}].
\eqn\magadj$$

\chapter{Discussion}

	Some of the most interesting applications of finite source effects in
microlensing involve the color changes due to differential limb darkening
(Witt 1995).  
For example, this effect can be exploited to measure the Einstein ring
size even when single-band photometric effects are undetectable, and it
is especially
useful in understanding planetary events (Loeb \& Sasselov 1995;
Gould \& Welch 1996).  The method given above cannot be directly applied
to limb darkened stars since constant surface brightness was assumed.  However,
one could model the source star as being composed of rings of constant
surface brightness, and each ring could be evaluated by taking the difference
of fluxes due to sources contained within two successive rings.

	The method given here is simpler than that of Bennett \& Rhie (1996)
in that it requires only a one-dimensional integral, but it is more complicated
in that one must find the individual image positions corresponding to the
source boundary.  (One must also find the parity and hence the magnification,
but this need be done only once for each image contour.)\ \
The method of choice therefore depends on the lens system.
For planetary system lenses, it is often possible to treat the effect of the
planet as
a perturbation on the background shear generated by its parent star.  In these
cases, the lens equation can be reduced to a quartic equation (Gould \& Loeb
1992) which can be solved analytically.  The method given here is therefore
far more efficient.  
In general, binary lenses require solution of a fifth
order equation.  There are standard packages that do this, but they require 
substantially longer computations than does the quartic case.  
Nevertheless, given that one is searching for images of points along the 
continuous
one dimensional boundary of the source, it should be possible to speed up
the fifth order programs by using the solution found for one point as a
trial solution for the next.  However, for more complicated lenses, 
two-dimensional
integration over the image plane may be preferable.

{\bf Acknowledgments} :  We would like to thank Scott Gaudi for stimulating
discussions and helpful comments.
This work was supported in part by grant AST 94-20746 from the 
NSF and in part by grant NAG5-2864 from NASA.
\smallskip
\singlespace
\Ref\alard{Alard, C.\ 1996, in Proc. IAU Symp.\ 173 (Eds.\ C.\ S.\ Kochanek, 
J.\ N.\ Hewitt), in press (Kluwer Academic Publishers)}
\Ref\ab{Alcock, C., et al.\ 1995, ApJ, 445, 133}
\Ref\ac{Alcock, C., et al.\ 1996a, ApJ, 461, 84}
\Ref\ac{Alcock, C., et al.\ 1996b, ApJ, submitted}
\Ref\ac{Alcock, C., et al.\ 1996c, ApJ, submitted}
\Ref\af{Aubourg, E., et al.\ 1995, A\&A, 301, 1}
\Ref\br{Bennett, D.\ \& Rhie, S.\ 1996, preprint}
\Ref\al{Gould, A.\ 1994, ApJ, 421, L75}
\Ref\al{Gould, A.\ 1996, PASP, 108, 465}
\Ref\gtwo{Gould, A., \& Loeb, A.\ 1992, ApJ, 396, 104}
\Ref\gtwo{Gould, A., \& Welch, R.\ L.\ 1996, ApJ, 464, 212}
\Ref\Lio{Liouville, J.\ 1837, Journal de Math\'ematiques Pures et 
Appliqu\'ees, 2, 16}
\Ref\ls{Loeb, A.\ \& Sasselov, D.\ 1995, ApJ, 449, L33}
\Ref\nem{Nemiroff, R.\ J.\ \& Wickramasinghe, W.\ A.\ D.\ T.\ 1994, ApJ, 424, 
L21}
\Ref\Pratt{Pratt, M., et al.\ 1996, in Proc. IAU Symp.\ 173 
(Eds.\ C.\ S.\ Kochanek, 
J.\ N.\ Hewitt), in press (Kluwer Academic Publishers)}
\Ref\oglea{Udalski, A., et al.\ 
1994a, Acta Astronomica 44, 165}
\Ref\oglec{Udalski, A., et al.\ 
1994b, ApJ 436, L103} 
\Ref\witt{Witt, H.\ 1995, ApJ, 449, 42}
\Ref\wm{Witt, H., \& Mao, S.\ 1994, ApJ, 430, 505}

\refout
\end